\documentclass[a4paper,fleqn,usenatbib]{mnras}

\usepackage{mathptmx}

\usepackage[T1]{fontenc}
\usepackage{ae,aecompl}

\usepackage{graphicx}	
\usepackage{amsmath}	
\usepackage{amssymb}	
\usepackage{color}
\title[The short-timescale evolution of QPO phase lags]{Probing the origin of quasi-periodic oscillations: the short-timescale evolution of phase lags in GRS 1915+105}

\author[J. van den Eijnden et al.]{
Jakob van den Eijnden,\thanks{E-mail: a.j.vandeneijnden@uva.nl}
Adam Ingram and Phil Uttley
\\
Anton Pannekoek Institute for Astronomy, University of Amsterdam, Science Park 904, 1098 XH Amsterdam, The Netherlands
}

\date{Accepted XXX. Received YYY; in original form ZZZ}

\pubyear{2016}

\begin{document}
\label{firstpage}
\pagerange{\pageref{firstpage}--\pageref{lastpage}}
\maketitle

\begin{abstract}

\noindent We present a model-independent analysis of the short-timescale energy dependence of low frequency quasi-periodic oscillations (QPOs) in the X-ray flux of GRS 1915+105. The QPO frequency in this source has previously been observed to depend on photon energy, with the frequency increasing with energy for observations with a high ($\gtrsim 2$ Hz) QPO frequency, and decreasing with energy for observations with a low ($\lesssim 2$ Hz) QPO frequency. As this observed energy dependence is currently unexplained, we investigate if it is intrinsic to the QPO mechanism by tracking phase lags on (sub)second timescales. We find that the phase lag between two broad energy bands systematically increases for $5$ - $10$ QPO cycles, after which the QPO becomes decoherent, the phase lag resets and the pattern repeats. This shows that the band with the higher QPO frequency is running away from the other band on short timescales, providing strong evidence that the energy dependence of the QPO frequency is intrinsic. We also find that the faster the QPO decoheres, the faster the phase lag increases, suggesting that the intrinsic frequency difference contributes to the decoherence of the QPO. We interpret our results within a simple geometric QPO model, where different radii in the inner accretion flow experience Lense-Thirring precession at different frequencies, causing the decoherence of the oscillation. By varying the spectral shape of the inner accretion flow as a function of radius, we are able to qualitatively explain the energy-dependent behaviour of both QPO frequency and phase lag.
\end{abstract}




\section{Introduction}

Accreting stellar-mass black holes in binary systems regularly display quasi-periodic oscillations (QPOs) in their X-ray flux with frequencies drifting from $\sim 0.1-10$ Hz \citep[e.g.][]{vanderklis89}. Three main components can be identified in the spectrum of these sources: disk blackbody emission, power law emission from the inner accretion flow, and a reflection spectrum from photons reflected off the disk \citep{done07}. So-called 'Type-C' low-frequency QPOs \citep{casella05} are believed to originate from the inner accretion flow/corona that is associated with the Comptonized power-law component of the X-ray spectrum, as this component shows a much larger variability amplitude than the blackbody disk component \citep{sobolewska06, alexsson13}. Since currently no consensus on the origin of QPOs exists, we can generally divide QPO models into two broad categories: \textit{geometric} and \textit{intrinsic} models. In the former, the X-ray emission is constant but an oscillating accretion geometry quasi-periodically alters the observed flux. A possible origin for these geometric oscillations could be Lense-Thirring precession of the Comptonizing medium, due to misalignment of the black hole spin and the binary orbit \citep{stella98, stella99, ingram09}. Alternatively, in intrinsic models the emitted luminosity itself varies, for example due to changes in mass accretion rate \citep{tagger99, cabanac10} or due to a standing shock in the accretion flow \citep{chakrabarti93}. Recently, \citet{heil15} and \citet{motta15} confirmed that the QPO amplitude depends on the inclination of the binary orbit, strongly suggesting a geometric origin \citep{schnittman06b}. \citet{ingram15} found that the iron line equivalent width changes over a QPO cycle in GRS 1915+105, also strongly pointing towards a geometric origin. 

GRS 1915+105 is a galactic low-mass black hole binary (BHB) located at a {distance of $8.6^{+2.0}_{-1.6}$ kpc \citep{reid14}}, that was discovered in 1992 by \citet{castrotirado92}. It shows a wide variety of variability properties \citep{belloni97a, belloni97b}, which are difficult to interpret within the standard picture of BHB accretion states (\citealt{belloni10}, see \citealt{vanoers10} for a spectral comparison). \citet{belloni00} report the presence of 12 accretion classes based on the properties of its lightcurve and color-color diagrams, which can be interpreted as transition between three main states: a hard state (C) where the disk is truncated, and two soft states (A \& B, with a low and high flux respectively) where the inner disk extends further inwards. However, all three states show similarities to the canonical very-high state of BHBs \citep{reig03}. GRS 1915+105 shows QPOs in various frequency ranges in several of its classes \citep{morgan97}. In this paper, we will focus on the low-frequency QPO with frequencies between $\sim 0.5-10$ Hz, which occurs only in the (hard) C state and is probably equivalent to Type-C QPOs in other sources \citep{casella05}.  

Recently, the QPO frequency has been found to change with observed energy band in several BHBs. The energy dependence of the QPO in GRS 1915+105 has been studied by \citet{qu10} and \citet{yan12}. \citet{yan12} analysed all RXTE observation of GRS 1915+105 up to 2010, and found a smooth evolution of the dependence of QPO frequency on photon energy. For observations with a low QPO frequency ($\sim 0.4-2.0$ Hz) in the full energy band, the QPO frequency decreases with energy, while for observations with a high QPO frequency ($\sim 2.0-8.0$ Hz) in the full energy band, the QPO frequency increases with energy. Similarly, \citet{li13} found an increase in QPO frequency with energy in XTE J1550-564 for frequencies above $\sim 3.3$ Hz. However, below $\sim 3.3$ Hz no variations with energy were observed. \citet{li13b} reported comparable behaviour in the QPO in H1743-322, also exclusively showing frequency increases with photon energy. GRS 1915+105 is thus the only source to systematically show decreases of QPO frequency with energy. As this decrease is particularly surprising within the standard picture of accreting compact objects, we have chosen GRS 1915+105 for our further analysis of the energy dependence.  

Similarly, the QPO phase lag is known to be energy dependent. \citet{pahari13}, \citet{qu10} and \citet{reig00} all found a smooth relation between this phase lag and energy in GRS 1915+105. The same behaviour is present in XTE J1550-564 \citep{wijnands99}. For GRS 1915+105, \citet{qu10} show that the slope of this energy dependence changes systematically from positive for observations with low QPO frequency (i.e. hard photons lag soft photons) to negative for observations with high frequency (i.e. soft photons lag hard photons). Futhermore, by comparing multiple observations, \citet{qu10} and \citet{pahari13} also show that the phase lag between hard and soft photons decreases approximately log-linearly as a function of QPO frequency in the full band, switching from a hard to soft lag around $\sim 2$ Hz. This frequency of $\sim 2$ Hz is of particular interest; in observations with this QPO frequency, also no energy dependence of the QPO frequency and zero phase lag is observed. As such, this frequency sets the border between the low and high QPO frequencies.  

These recent results on the energy dependence of both the QPO frequency and phase lags pose several challenges for current QPO models: not all models predict an energy-dependent QPO frequency, and none can account for the decrease of frequency with energy observed in GRS 1915+105 \citep{qu10}. Intuitively, both a higher frequency and harder spectrum can be associated with a smaller radius in the accretion flow. Thus an increase of QPO frequency with energy could be expected, but a decrease is counterintuitive. Furthermore, no explanation exists for the energy and frequency dependence of the phase lags. Finally, the link between all three is not fully understood. Hence, the simplest explanation is that the QPO mechanism possesses only a single oscillation frequency. The observed relation between QPO frequency and energy could simply arise if the hardness of the QPO lightcurve correlates with the QPO frequency. In that interpretation, both the hard and soft band always show the same frequency jitter, but the changing hardness weights the QPO frequency differently in different energy bands. This would lead to observed differences in QPO frequency between different energy bands, even though there is only a single underlying QPO frequency. 

In this paper, we test this hypothesis that the observed energy dependence of the QPO frequency arises due to hardness-frequency correlations in the QPO lightcurve. We have developed a novel, model-independent approach to investigate properties of the QPO, such as hardness, frequency and phase lag, on the timescale of single QPO cycles, by removing non-QPO variability from the observed lightcurves. This allows us to test for biases causing the observed energy dependence, as explained above, by tracking QPO frequencies and hardness on short timescales. Unexpectedly, we find strong evidence that the observed frequency differences are a genuine property of the underlying QPO mechanism. We also find that the phase lag at the QPO frequency increases systemically on the timescale of $5-10$ QPO cycles as a result of this frequency difference, before resetting once the QPO has become decoherent. We interpret our results in a geometric toy model where the innermost accretion flow is subject to differential precession. By varying the shape of the emitted X-ray spectrum as a function of radius, we are able to qualitatively explain the observed energy dependencies of the QPO properties.

\section{Observations and timing analysis}
\label{method}
\begin{figure}
  \begin{center}
    \includegraphics[width=\columnwidth]{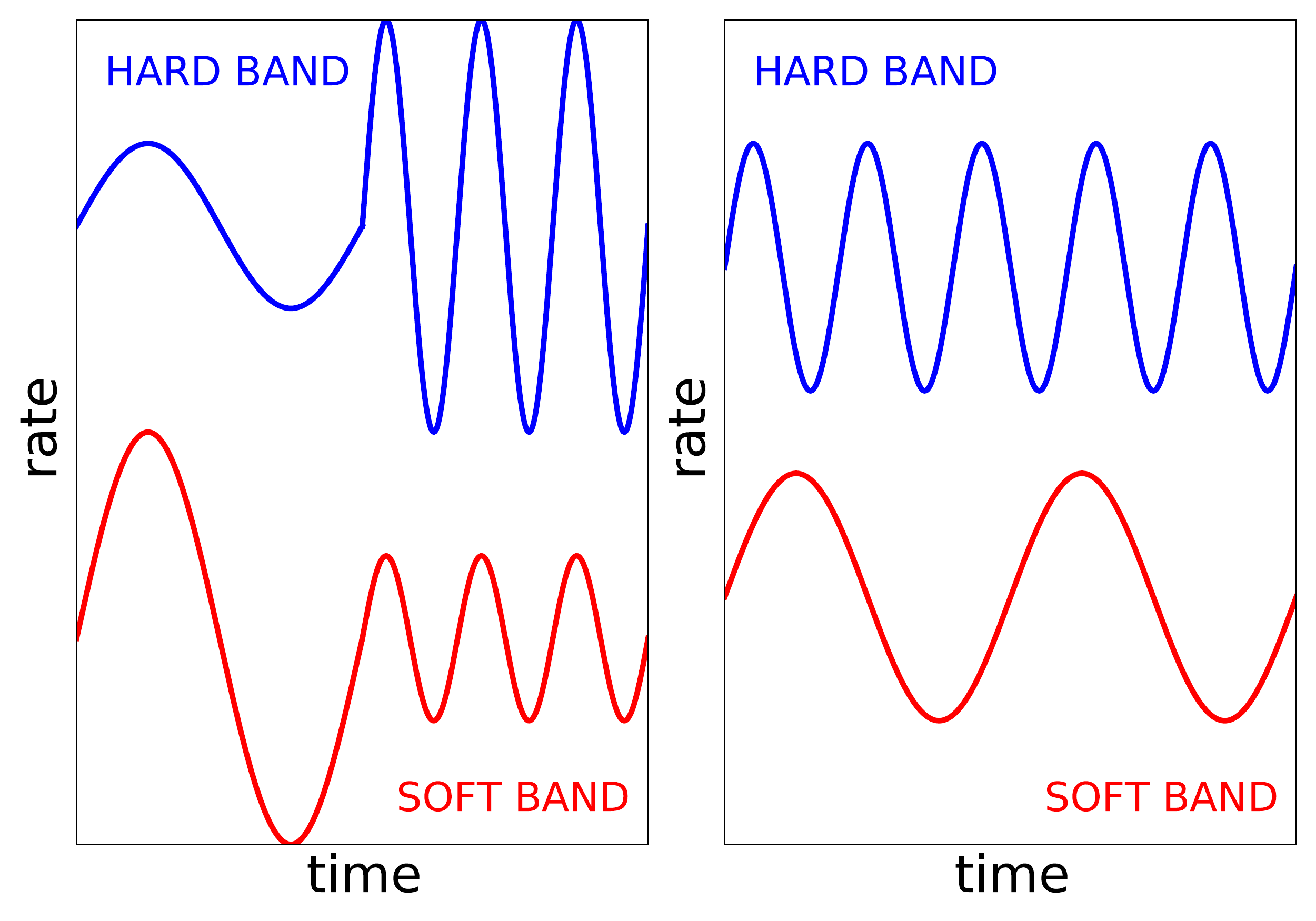}
    \caption{Cartoon depiction of the two considered scenarios for the observed energy dependence of the QPO frequency. In the scenario on the left, the frequencies in both energy bands are always the same, while in the scenario on the right, the frequencies are different. For clarity, the differences in frequency are exaggerated compared to actual observed frequency differences (listed in Table \ref{tab:obsoverview}).}
    \label{fig:2}
  \end{center}
\end{figure}

\noindent In this paper, we consider two plausibel origins for the energy dependence of the QPO frequency, which are depicted schematically in Figure \ref{fig:2}. In the left scenario, the QPO lightcurves in the two energy bands always have the same frequency. However, this frequency changes as a function of time. Whenever the frequency is high, the hard band lightcurve has a large amplitude compared to the soft band lightcurve (where amplitude refers to the maximum deviation from the mean and not the rms amplitude). When the frequency is low, the amplitudes are reversed, i.e. the amplitude is higher in the soft band. In this scenario, the power spectra in the two energy bands would show a QPO frequency weighted towards the high amplitude segments of the lightcurve. Thus the power spectra would show a different QPO frequency, even though the frequencies are always the same. In the alternative scenario, on the right, the QPO lightcurves in the two energy bands simply posses a different frequency. While this might seem to be a simpler explanation of the observed energy-dependence of the QPO frequency, the former scenario is more consistent with current models as there is only a single QPO frequency. Furthermore, in the latter scenario, the different QPO frequencies would cause a runaway between different energy bands over long timescales, which is contradicted by the coherent nature of the QPO. 

There are two tests to distinguish between these two possible explanations: first, in the left scenario in Figure 1, the amplitude of the QPO lightcurve should be either correlated with the frequency in the hard band and anticorrelated with frequency in the soft band, or vice versa. In the right scenario, such (anti)correlations are not necessarily expected. Secondly, if the QPO frequency in both energy bands is always the same, the phase lag is expected to stay constant. However, if both energy bands posses a different QPO frequency, this phase lag would systematically change over time. As we know that the QPO is coherent on long timescales, these changes in phase lag would occur only on very short timescales. 

We have developed a model-independent method to search both for correlations between QPO frequency and amplitude, and for short-timescale variations in phase lag. The method consists of broadly four steps: we (1) extract light curves in two broad energy bands and calculate their power spectra, (2) filter these light curves in order to conserve only the QPOs, (3) determine the frequency and amplitude of each QPO cycle, and (4) track the phase lag between the energy bands on the timescale of individual QPO cycles. Steps (1), (2) and (3) are desribed in section \ref{sec:filter}, while step (4) is described in section \ref{sec:phaselags}.

\begin{table*}
 \begin{center}
  \caption{\small{Overview of analysed \textit{RXTE} observations. Listed are the ObsId, observation date, fundamental QPO frequency in the reference band $\nu^{\rm ref}_{0}$, the difference in QPO frequency (hard - soft) $\Delta \nu_0$, the Q-factor in the full energy band ($\nu_{\rm QPO}/\text{FWHM}$), the used energy bands (I: $1.94-6.89$ and $6.89-12.99$ keV; II: $1.94-6.54$ and $6.54-12.99$ keV; III: $2.13-6.72$ and $6.72-12.63$ keV), the reduced $\chi^2$ of the power spectral fit $\chi^2/\text{d.o.f.}$, and the slope and offset of the phase lag evolution (see Section \ref{sec:pl}).}}
  \label{tab:obsoverview}
 
   \begin{tabular}{lccccclcc}
  \hline
  ObsID & Date & $\nu^{\rm ref}_{0}$ (Hz) & $\Delta \nu$ (Hz)& Q-factor & Energy Bands&$\chi^2/\text{d.o.f.}$ & Slope & Offset \\
  \hline \hline
  10258-01-06-00a & 29-08-1996 & $5.141 \pm 0.024$ & $0.078 \pm 0.036$ & 5.9 & I & $1.02$ & $-0.20$ & $-0.06$  \\
  10408-01-21-02 & 07-07-1996 & $8.107 \pm 0.043$ &  $0.313 \pm 0.071$ & 6.2 & I & $1.11$ & $-0.40$ & $0.01$ \\
  10408-01-22-00 & 11-07-1996 & $3.480 \pm 0.004$ &  $ 0.008 \pm 0.008 $ & 8.5 & I & $1.22$ & $-0.08 $ & $-0.06 $ \\ 
  10408-01-22-01 & 11-07-1996 & $2.777 \pm 0.005$ &  $ 0.003\pm 0.007$ & 6.8 &I &$1.24 $ & $-0.03 $ & $-0.03 $ \\
  10408-01-22-02 & 11-07-1996 & $2.560 \pm 0.004$ &  $0.000 \pm 0.007 $ & 6.3 &I &$1.36$ & $-0.04 $ & $-0.02 $ \\
  10408-01-27-00 & 26-07-1996 & $0.632 \pm 0.002$ &  $-0.003 \pm 0.002$ & 5.3 &I &$1.08$ & $0.08 $ & $0.13 $ \\
  10408-01-28-00 & 03-08-1996 & $0.966 \pm 0.002$ &  $-0.005 \pm 0.003$ & 5.3 &I &$1.34$ & $0.04 $ & $0.10 $ \\
  10408-01-29-00a & 10-08-1996 & $1.658 \pm 0.004$ & $-0.004 \pm 0.006$ & 10.8 &I &$1.03$ & $-0.01 $ & $0.06 $ \\
  10408-01-29-00b & 10-08-1996 & $1.856 \pm 0.004$ & $ 0.006 \pm 0.005 $ & 9.9&I&$1.18$ & $-0.07 $ & $0.07 $ \\
  10408-01-29-00c & 10-08-1996 & $1.963 \pm 0.004$ & $-0.010 \pm 0.006$ & 6.8 &I &$1.40$ & $-0.03 $ & $0.02 $ \\
  10408-01-30-00 & 18-08-1996 & $4.944 \pm 0.010$ &   $0.041 \pm 0.017$ & 3.0 &I &$1.07$ & $-0.05 $ & $-0.11 $ \\
  10408-01-31-00a & 25-08-1996 & $ 4.084\pm 0.008$ & $0.013 \pm 0.01$ & 9.5 & I&$0.96 $ & $-0.05 $ & $-0.09 $ \\
  10408-01-31-00b & 25-08-1996 & $4.439 \pm 0.009$ & $ 0.04\pm 0.014$ & 5.0 &I &$1.17$ & $-0.11 $ & $-0.07 $ \\
  10408-01-31-00c & 25-08-1996 & $3.514 \pm 0.006$ & $0.008 \pm 0.009$ & 6.7 & I&$1.25$ & $-0.0 $ & $-0.08 $ \\
  10408-01-32-00 & 31-08-1996 & $6.121 \pm 0.018$ &  $0.296 \pm 0.029$ & 3.9 &I &$1.38$ & $-0.39 $ & $-0.02 $ \\ \hline
  20402-01-48-00 & 29-09-1997 & $7.639 \pm 0.034$ &  $0.242 \pm 0.054$ & 5.9 &II &$1.35$ & $-0.43 $ & $-0.02 $ \\
  20402-01-50-01 & 16-10-1997 & $1.042 \pm 0.003$ &  $-0.005 \pm 0.004$ & 6.1 &II &$1.11$ & $0.03 $ & $0.10 $ \\
  30182-01-01-00 & 08-07-1998 & $1.870 \pm 0.009$ &  $0.007 \pm 0.013$ & 8.1 &II &$1.08$ & $-0.08 $ & $0.04 $ \\
  30402-01-11-00a & 20-04-1998 & $5.245 \pm 0.032$ & $0.127 \pm 0.033$ & 7.3 & II&$1.09$ & $-0.33 $ & $-0.01 $ \\
  30402-01-11-00b & 20-04-1998 & $5.857 \pm 0.017$ & $0.129 \pm 0.039$ & 3.6 &II &$1.23$ & $-0.24 $ & $-0.06 $ \\
  30703-01-20-00 & 24-05-1998 & $0.696 \pm 0.002$ &  $-0.004 \pm 0.003$ & 5.3 &II &$1.01$ & $0.03 $ & $0.14 $ \\
  30703-01-35-00 & 25-09-1998 & $2.464 \pm 0.006$ &  $0.008 \pm 0.009$ & 5.9 &II &$1.14$ & $-0.03 $ & $-0.05 $ \\ \hline
  40703-01-38-01 & 15-11-1999 & $7.114 \pm 0.030$ &  $0.373 \pm 0.044$ & 4.1 &III &$1.09$ & $-0.57 $ & $0.09 $ \\
  40703-01-38-02 & 15-11-1999 & $7.943 \pm 0.032$ &  $0.282 \pm 0.048$ & 7.2 &III &$1.27$ & $-0.89 $ & $0.22 $ \\
  
  \hline  
  \end{tabular}
  \end{center}
\end{table*}

\subsection{Data reduction and optimal filtering}
\label{sec:filter}

For our analysis, we select 24 \textit{Rossi X-ray Timing Explorer} (RXTE) Proportional Counter Array (PCA) observations of GRS 1915+105, based on the observations discussed in \citet{qu10}, \citet{pahari13} and \citet{yan13}. 
The observations are selected to evenly span a range in QPO frequency from $\sim 0.5$ to $8$ Hz. Table \ref{tab:obsoverview} summarizes the main properties of these observations. Using the standard FTOOLS\footnote{\href{https://heasarc.gsfc.nasa.gov/ftools/ftools\_menu.html}{https://heasarc.gsfc.nasa.gov/ftools/ftools\_menu.html}} package, 
we extract binned mode data to produce light curves in three energy bands: a soft band from $\sim2$ to $\sim6.7$ keV, a hard band from $\sim6.7$ to $\sim13$ keV, and a reference band covering both energy ranges. 
Due to changes in the PCA gain, the exact energy bands differ slightly between observations. The exact energy bands are indicated in Table \ref{tab:obsoverview} for all observations. We extract all observations using a $1/128$ s time resolution, which yields a Nyquist frequency of $64$ Hz for the subsequent analysis.

We divide all lightcurves into $8$ second segments and for each one calculate the power density spectrum (PDS) with a $1/8$ Hz resolution. After applying the rms-squared normalisation \citep{belloni90} we 
average the separate power spectra into one PDS per lightcurve to reduce the standard errors. Using \textsc{XSPEC} v12\footnote{\href{https://heasarc.gsfc.nasa.gov/xanadu/xspec/}{https://heasarc.gsfc.nasa.gov/xanadu/xspec/}}, 
we fit the average power spectra with a model consisting of a constant white noise, two broad band noise (BBN) Lorentzians with a fixed centroid frequency of $0$ Hz, and a Lorentzian for the QPO fundamental and each (sub)harmonic. 
We fit all energy bands within the same observation separately with the same model, as linking parameters between energy bands generally results in worse fits. 
Details of the resulting fits, including reduced $\chi^2$ values and QPO frequencies, 
are listed in Table \ref{tab:obsoverview}. The errors shown are the one sigma confidence intervals.

In order to study the behaviour of only the QPO, and remove broad band and Poisson noise contributions to the variability, we apply an optimal filtering technique based on the method in \citet{filtering}. We assume that the observed count
rate $c(t)$ consists of the true QPO signal $q(t)$ and an added noise component $n(t)$:
\begin{equation}
 c(t) = q(t) + n(t)
 \label{eg:counts}
\end{equation}
Our aim is to remove $n(t)$ in order to estimate $q(t)$ as accurately as possible. The optimal filter provides such an estimate of the true QPO signal, $\tilde{q}(t)$, by minimizing the squared difference between $q(t)$ and $\tilde{q}(t)$. In practice, the filter $F(\nu)$ is applied by multiplication with the Fourier transform of 
the count rate (capitalized variables indicate the Fourier transform):
\begin{equation}
 \tilde{Q}(\nu) = F(\nu)\cdot C(\nu)
 \label{eq:filter1}
\end{equation}
The filter in Fourier space is given by
\begin{equation}
 F(\nu) = \frac{|Q(\nu)|^2}{|C(\nu)|^2}
 \label{eq:filter2}
\end{equation}
and thus requires an estimate of the actual QPO power spectrum $|Q(\nu)|^2$, for which we apply the fitted QPO Lorentzian. The remaining time series $\tilde{q}(t)$ estimates the true QPO lightcurve, without other variability contributions.

The optimal filter assumes that the QPO signal $q(t)$ and the noise contribution $n(t)$ are uncorrelated. In our method, the noise consists of 
the white noise, the BBN and any (sub)harmonic QPO peaks. Since the (sub)harmonics are clearly related to the fundamental QPO and correlations between the BBN and the QPO are known to exist \citep{heil11}, the assumption of uncorrelated noise does
not fully hold. This will especially spoil the filter at low frequencies, where the BBN is dominant, and at the (sub)harmonic frequencies. To cancel these effects, we apply an extra cut that removes all high and low frequencies outside the range $\nu_{\rm QPO} \pm \rm FWHM$, where $\nu_{\rm QPO}$ is the fitted QPO frequency in the considered energy band. As this does not remove the correlations at the QPO frequency, the filter remains slightly less then optimal. Alternative filters, that do not make assumptions about noise correlations, exist: for example, the tophat filter simply removes all power outside a certain frequency range. These filters are less accurate than the optimal filter and do not use any known properties of the QPO peak. For this reason, we apply the optimal filter for the subsequent analysis. However, our main results, presented in the next section, do not differ significantly when using the tophat filter. 

Both the optimal and alternative filters only affect the amplitude of the power spectrum, while leaving the phases unaltered. This implies that we can use the filtered lightcurves to measure phase lags in the subsequent analysis. However, this also means that while the BBN amplitude is removed, its phase lags are still present in the filtered lightcurve. This requires us the make the assumption that, at $\nu_{\rm QPO}$, the phase lags are dominated by the QPO and the contribution of the BBN is neglegible. We will discuss the effects of our choice of filter and the validity of this assumption in section \ref{sec:robust}.

\begin{figure}
  \begin{center}
    \includegraphics[width=\columnwidth]{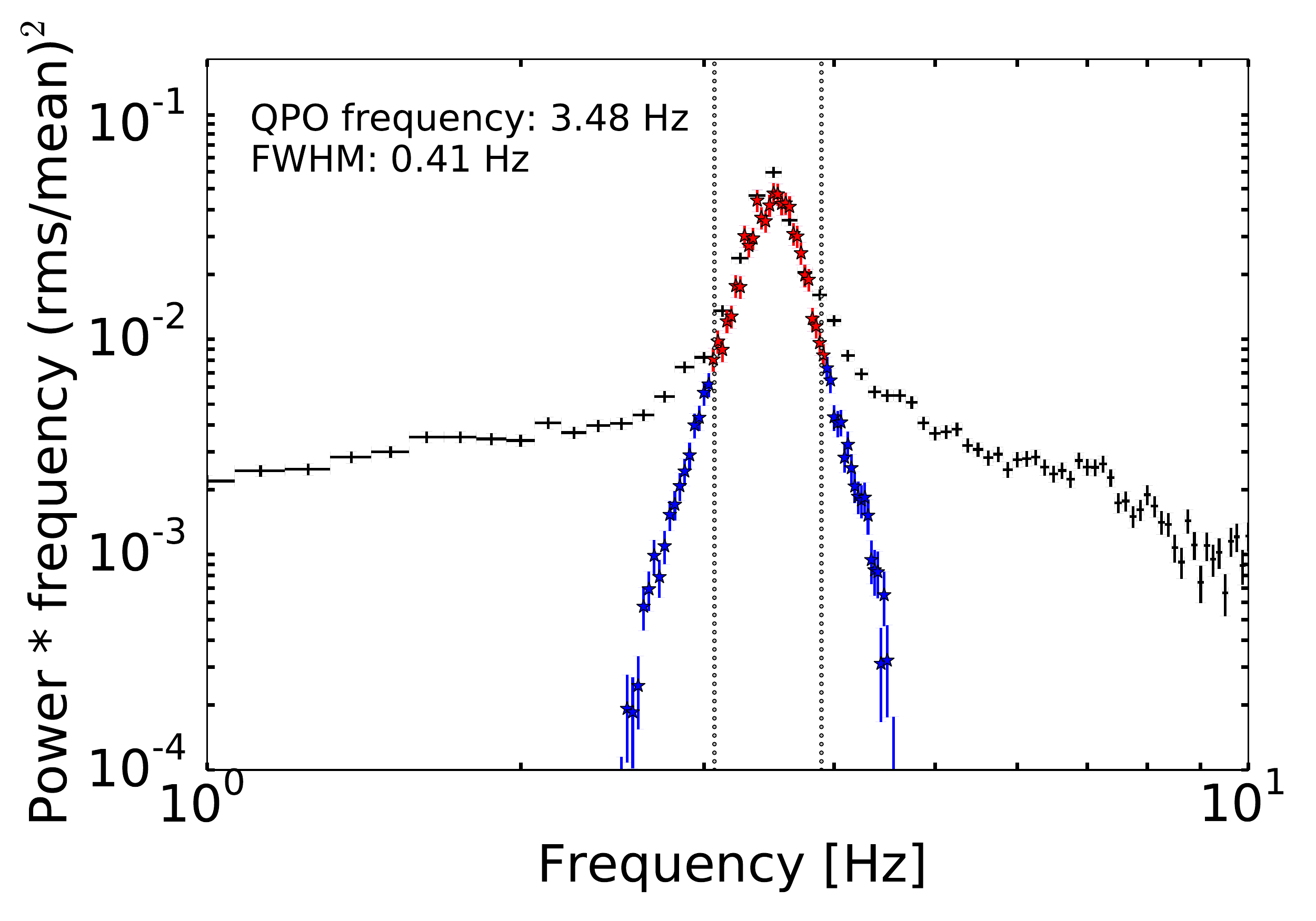}
    \caption{Explanatory example of the optimal filter in the frequency domain. The black points correspond to the observed power spectrum, the red and blue stars to the filtered power spectrum. We only use the inner (red) part of the filtered
power spectrum, within the range shown by the dotted lines ($\nu_{\rm QPO} \pm \rm FWHM$), to produce QPO light curves.}
    \label{fig:ps}
  \end{center}
\end{figure}

\begin{figure}
  \begin{center}
    \includegraphics[width=\columnwidth]{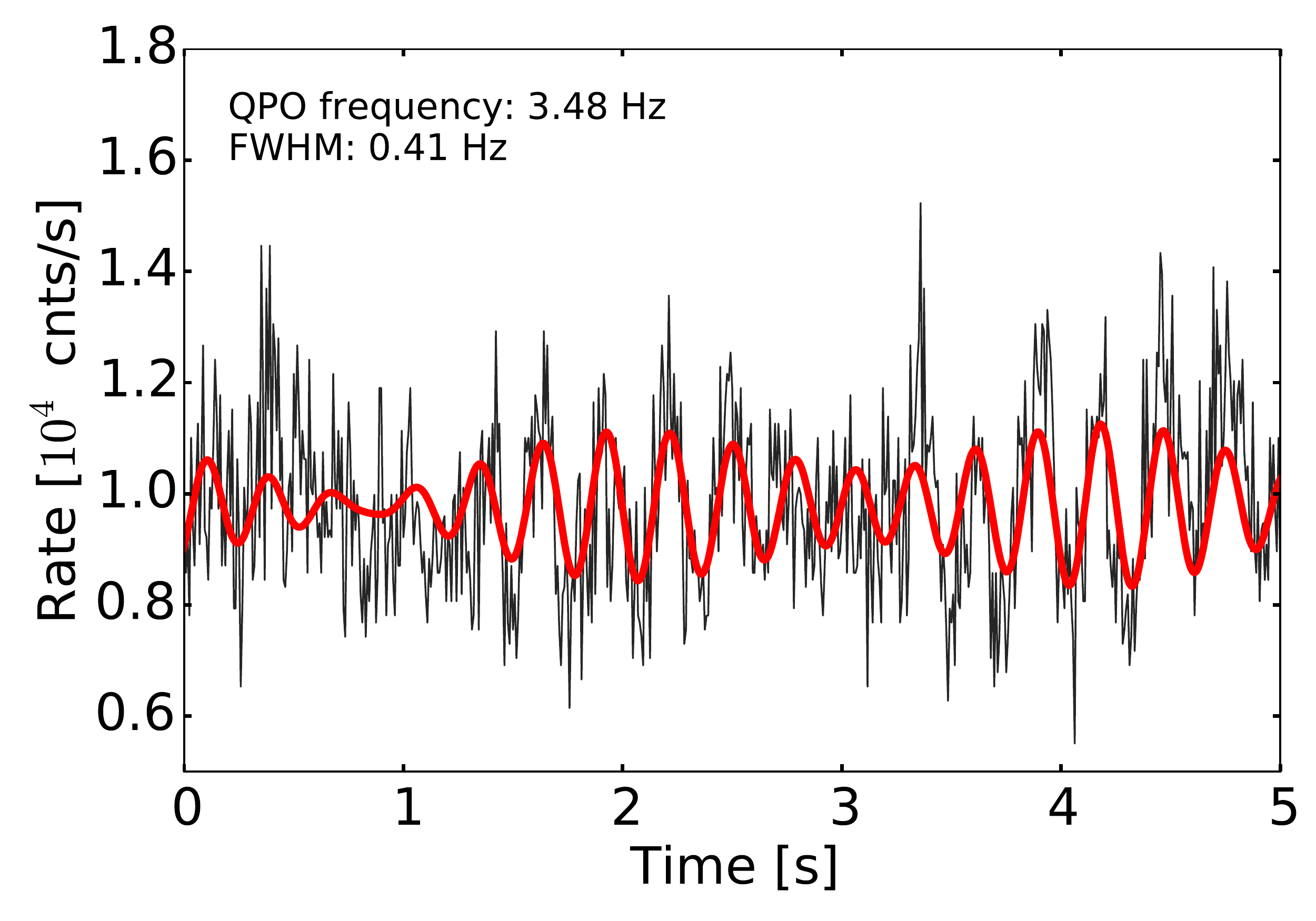}
    \caption{Example of the optimal filter in the time domain. The black and red curves are the light curves corresponding to respectively the black and red power spectra in Figure \ref{fig:ps}. The filtered light curve (red) clearly picks out the QPO,
while removing other variability present in the observation.}
    \label{fig:ls}   
  \end{center}
\end{figure}

Figure \ref{fig:ps} shows an example of an unfiltered and filtered power spectrum. The peak of the QPO is clearly sampled by the filtered power spectrum, while the power becomes zero outside the allowed frequency range. We use $64$ second segments of the light curves to produce the power spectra that are filtered, causing the difference in frequency resolution in Figure \ref{fig:ps}. It is possible to select longer segments since the power spectra are already 
fitted, so there is no need to average many power spectra to reduce standard errors. Figure \ref{fig:ls} shows the light curves corresponding to the power spectra in Figure \ref{fig:ps} in the same colors. As intended, the filtered light curve tracks the large overall oscillations, but does not sample the added noise contributions. 

In order to track each QPO cycle individually, we use simple linear interpolation to estimate the mean-crossings and extrema of the filtered reference band light curves. Defining a QPO cycle as a light curve segment including either three consecutive mean-crossings or two 
consecutive maxima, we can determine both the maximum amplitude and frequency of each individual cycle. This allows us to perform the first test of the energy-dependence of the QPO frequency: the aforementioned presence of (anti)correlations between amplitude and frequency. 

\subsection{Phase lags}
\label{sec:phaselags}

\begin{figure}
  \begin{center}
    \includegraphics[width=\columnwidth]{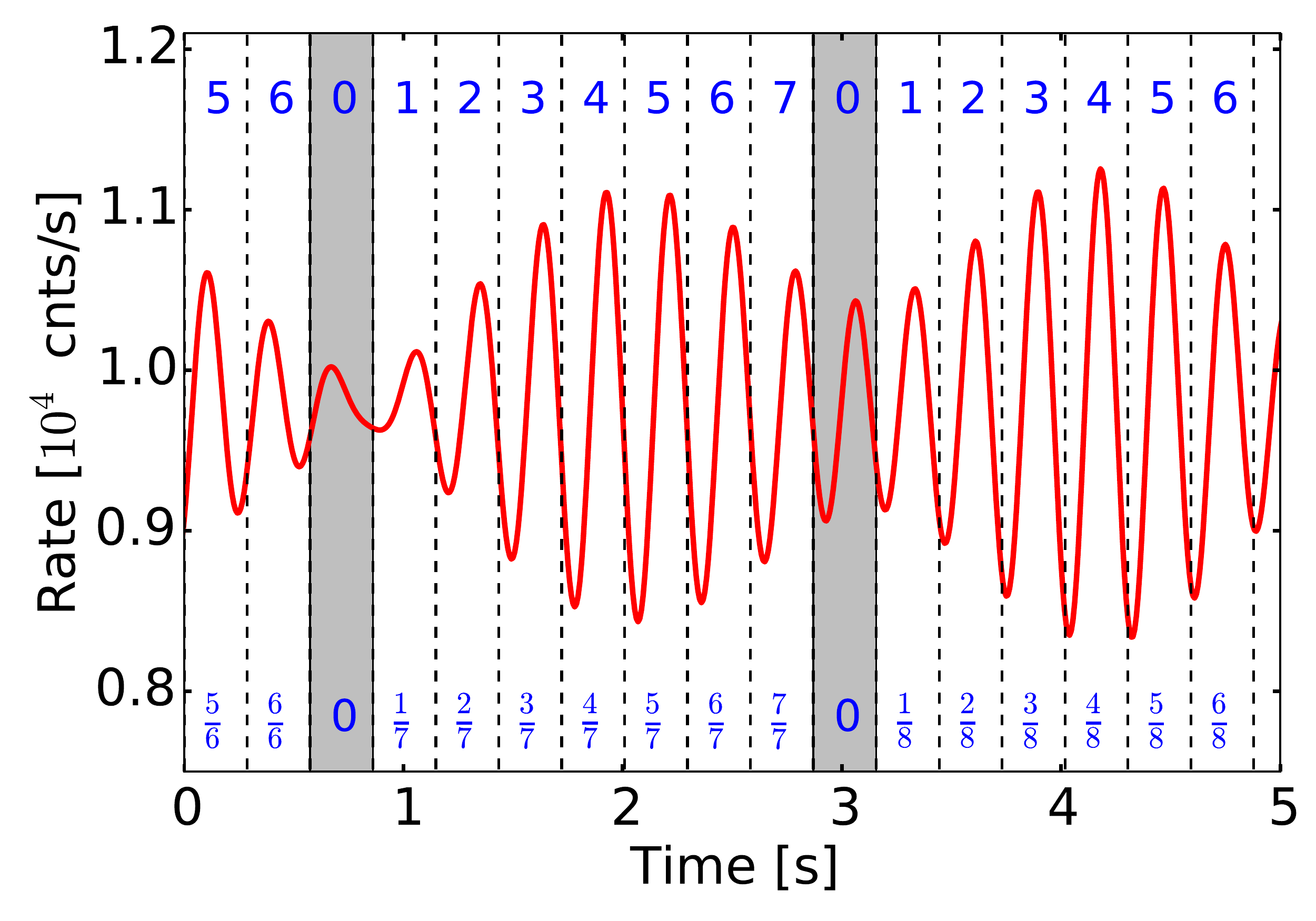}
    \caption{Categorizing method for the position of segments in the filtered lightcurve. The dotted lines indicate the edges of each segment. Segments containing the start of a coherent interval, equivalent to a minimum
QPO amplitude, are indicated by a grey band. Segments are grouped together based on either absolute label (upper) or ratio of label to full coherent interval length (lower).}
    \label{fig:coh}   
  \end{center}
\end{figure}

The second method to test the energy-dependence of the QPO frequency is to track the phase lag over time: if the different energy bands harbor a different QPO frequency, this phase lag should change systematically. But as was already stated, these changes should occur on short timescales only, since the QPO is coherent on longer timescales. Hence, we first need to identify the timescale on which to search for variations in phase lag. 

As is visible in Figure \ref{fig:ls}, the filtered lightcurves show that the QPO amplitude repeatedly rises and falls in an enveloping modulation. While the individual envelopes differ in length, the mean length of the envelopes in an observed lightcurve is tightly correlated with the Q-factor ($\nu_{\rm QPO}/\text{FWHM}$) in that lightcurve. In other words, the more coherent the QPO, the longer the envelopes in the filtered QPO lightcurve. Hence, we name these envelopes \textit{coherent intervals} and select these as the timescales within which we search for the changes in phase lag that might signal a genuine energy dependence. We designed our subsequent analysis such that it allows us to track the phase lag throughout these coherent intervals. As such, it consists of three independent steps: (1) splitting each observation in segments, (2) identifying the positions of these segments in coherent intervals, and (3) calculating the phase lag by combining lightcurve segments located at similar positions in their respective coherent interval. 

First, we divide the time-axis of each observation into short segments of length $1/\nu^{\rm ref}_{0}$, where $\nu^{\rm ref}_{0}$ is the QPO frequency in the full $2-13$ keV band (hereafter referred to as the \textit{full-band QPO frequency}). Using the optimally-filtered reference band lightcurve, we search for minima in QPO amplitude and identify these as the start and end points of coherent intervals. For the second step, we label each segment with a number based on its position inside its respective coherent interval, as illustrated in Figure \ref{fig:coh}: starting with zero for the segment containing the start of the coherent interval, we increase the label of each successive segment by one, up until the end of the coherent interval. We define the \textit{absolute position} as the label of a segment, and the \textit{fractional position} as the ratio of the label to 
the total amount of segments in the coherent interval. These positions are shown in Figure \ref{fig:coh} as the upper and lower numbers, respectively. As we only used the filtered QPO amplitudes in this process of defining segment labels, this first step is completely independent of the phase or frequency properties of the QPO. 

As the final step, we filter the hard and the soft band using their respective best fitting parameters. We normalise both lightcurves in each segment by subtracting their mean and dividing by their standard deviation, and cross correlate each set of simultaneous segments in the two energy bands. The position of the peak in the resulting cross correlation function (CCF) indicates the time lag between the two energy bands in that specific segment. In order to increase the signal-to-noise ratio in the CCFs, we average CCFs with either the same absolute or similar relative positions. In order to account for the uncertainty in our simple determination of the starts of the coherent intervals, we do not consider the zero-labeled segments in the subsequent analysis.

The time lag $\tau(\nu^{\rm ref}_{0})$ between the hard and soft band can then be measured as the location of the peak of the averaged CCF. Due to the extra cuts in the optimal filter and the frequency difference between the two energy bands, no analytical description of this
CCF exists. Therefore, we fit a skewed Gaussian model to the central peak of the averaged CCF. As the maximum of the skewed Gaussian model is not represented by a single parameter, we then apply bootstrapping to the averaging of the CCFs to determine the one-sigma uncertainty on the time lag \citep[see e.g.][section 15.6]{filtering}. Finally, we convert the time lags into phase lags $\phi(\nu^{\rm ref}_{0})$ using $\phi(\nu^{\rm ref}_{0}) = \tau(\nu^{\rm ref}_{0}) \cdot 2 \pi \nu^{\rm ref}_{0}$. Thus, we obtain an estimate of the average phase lag between the hard and soft band at different absolute or fractional positions in the coherent intervals. 

\section{Results}
\label{results}
\subsection{Test 1: frequency-amplitude correlations}

\begin{figure*}
  \begin{center}
    \includegraphics[width=\textwidth]{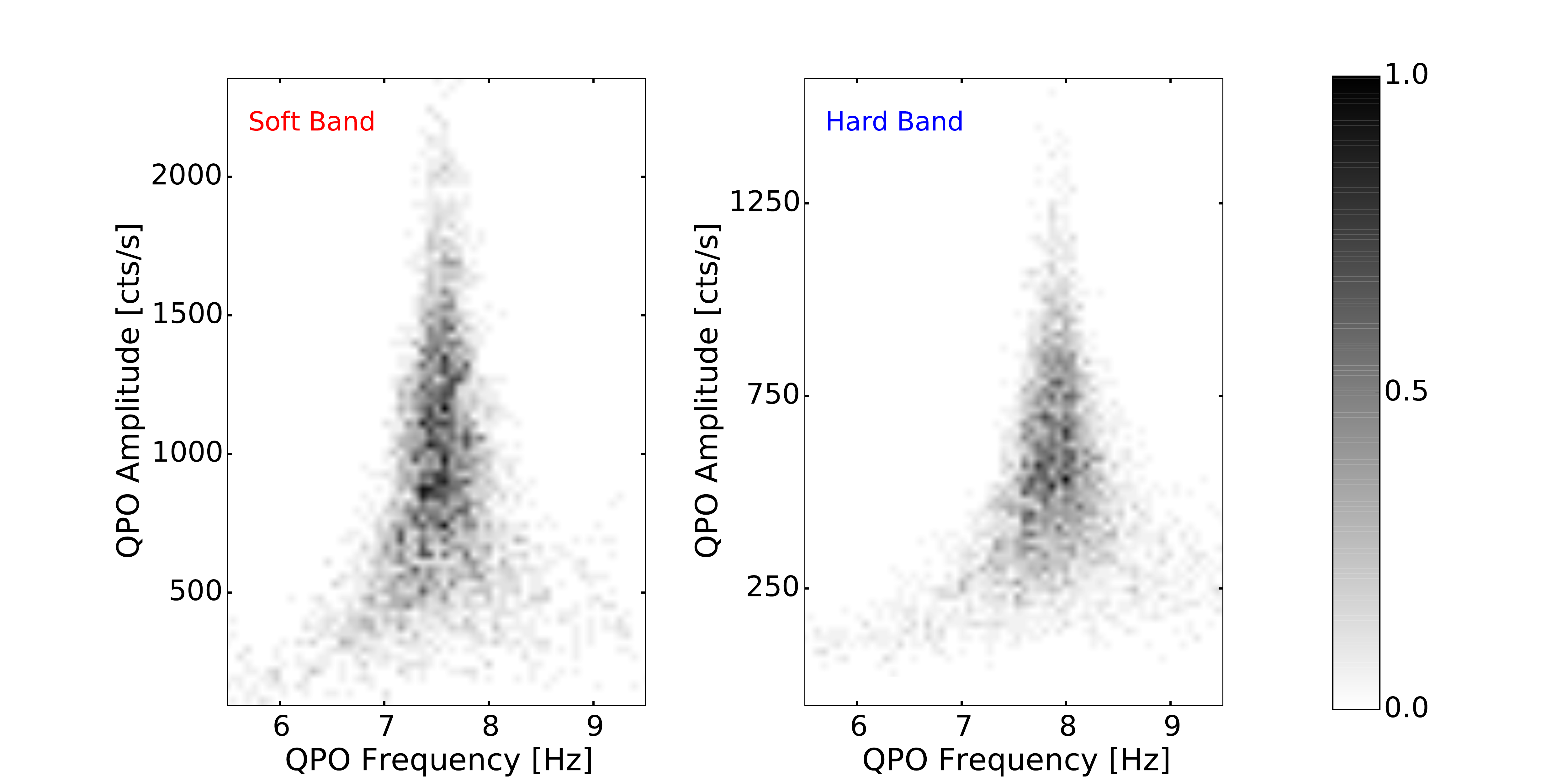}
    \caption{Representative example of the relation between amplitude and frequency of individual QPO cycles. The panels show a two-dimensional histogram of these amplitudes and frequencies, rescaled such that the highest counts are equal to one. The left and right panel show the relation for the soft and hard band, respectively.}
    \label{fig:corr}   
  \end{center}
\end{figure*}

In Figure \ref{fig:corr}, we show a representative example of the relation between the amplitude and frequency of single QPO cycles. The panels show a two-dimensional histogram of the amplitude and frequency of individual QPO cycles, rescaled such that the highest counts are equal to one. The left and right panel show the relation for the soft and hard band, respectively. Two immediate conclusions can be drawn from the Figure: first, both the soft and hard band show no apparent (anti)correlation between frequency and amplitude. Secondly, the panels of the energy bands appear strikingly similar, even though the center of the distribution lies at a different QPO frequency. This lack of correlation and similarity between the energy bands is present in all observations analysed. Thus, this first tests unexpectedly points towards different QPO frequencies in the different energy bands, i.e. a genuine energy dependence of the QPO frequency. 

As a test of our method, we converted the observed power spectra into maximally stochastic light curves using the method of \citet{timmer95}, i.e. the best fitting power spectra were used to generate simulated lightcurves with the properties of Gaussian noise. After applying the optimal filtering and the tracking of QPO properties, these simulated light curves show a completely similar relation between amplitude and frequency as the observed ones shown in Figure \ref{fig:corr}. This similarity indicates that the triangular shapes in the amplitude-frequency plane are simply an inherent property of a noise process with the observed power spectra. The triangular shape itself is likely due to fact that at lower amplitudes, the determination of individual QPO cycles becomes less precise due to noise contributions. 

\subsection{Test 2: phase lag evolution}
\label{sec:pl}
\begin{figure*}
  \begin{center}
    \includegraphics[width=\textwidth]{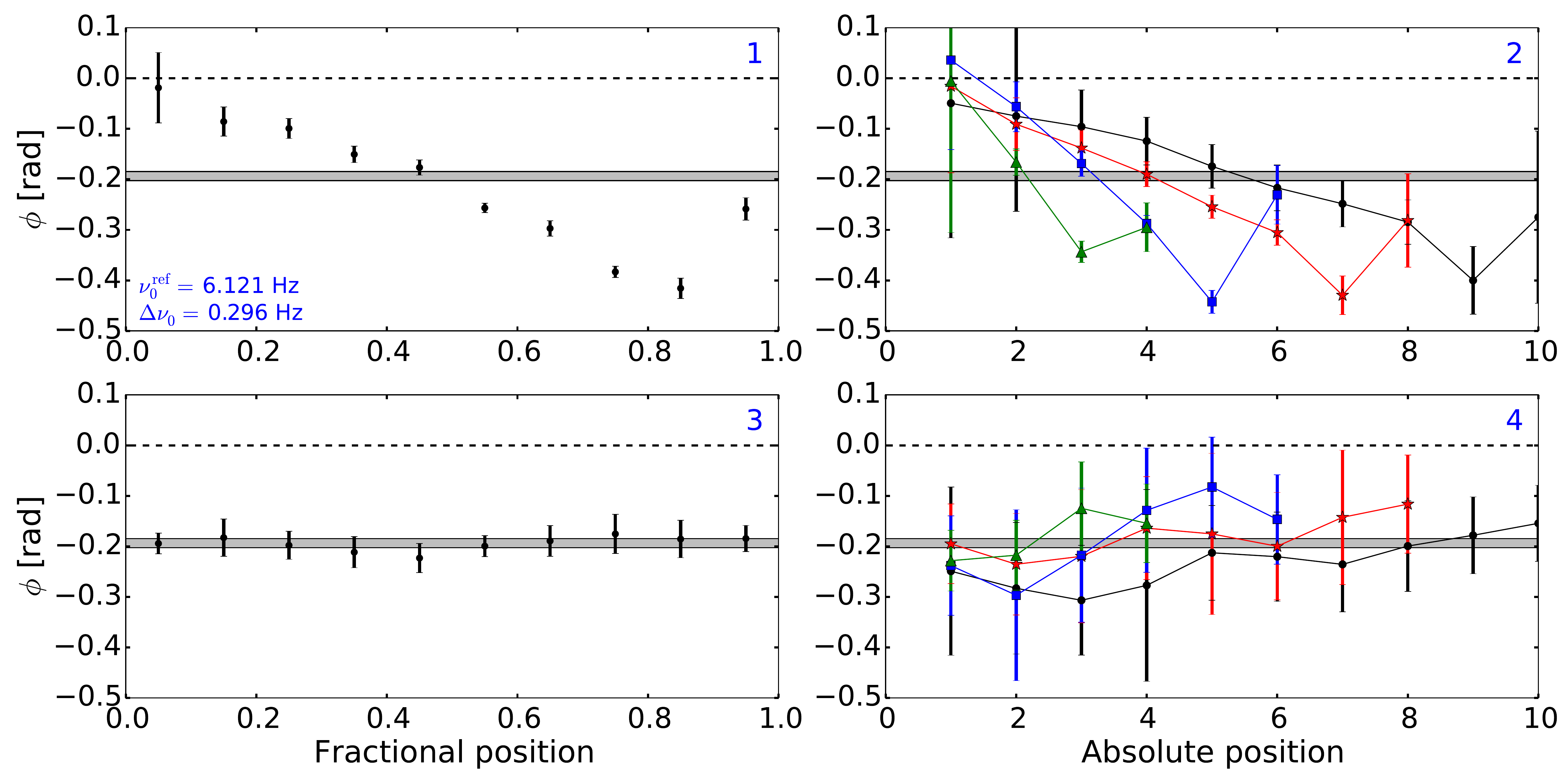}
    \caption{Evolution of the phase lag between the hard and soft QPO light curve throughout a coherent interval for a $\sim 6.1$ Hz QPO ($\Delta \nu_0 \sim 0.3$ Hz). Panels 1 and 2 show the phase lags versus fractional and absolute position in a coherent interval, respectively. The lag evidently increases and resets at the end. Panels 3 and 4 show the corresponding randomized checks of our method, and are expected to hover around the average phase lag, shown by the grey band. Positive values of $\phi$ indicate hard lags. See full text for details.}
    \label{fig:main_h}   
  \end{center}
\end{figure*}

\begin{figure*}
  \begin{center}
    \includegraphics[width=\textwidth]{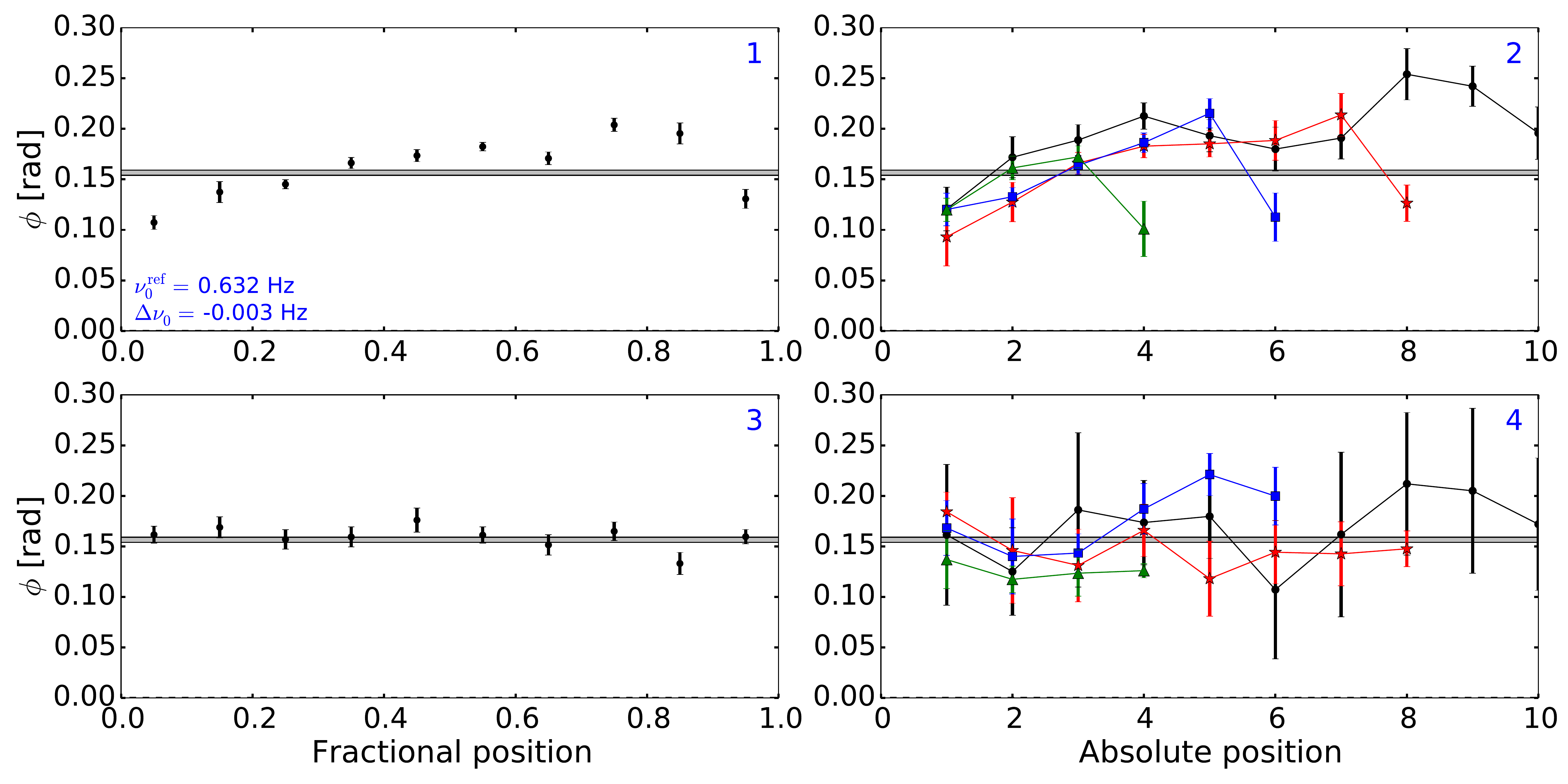}
    \caption{Evolution of the phase lag between the hard and soft QPO light curve throughout a coherent interval for a $\sim 0.6$ Hz QPO ($\Delta \nu_0 \sim -0.003$ Hz). Same as Figure \ref{fig:main_h}}
    \label{fig:main_l}   
  \end{center}
\end{figure*}

As a second test, we search for changes in the phase lag on short timescales. In panel 1  of Figure \ref{fig:main_h},
we show a representative example of the evolution of the phase lag $\phi$ as a function of fractional position within a coherent interval, with $\phi > 0$ corresponding to hard lags. The phase lag clearly increases from no lag at the start of the coherent intervals
(fractional position of $0.0$) to a significant soft lag at the end (fractional position of $1.0$). This shows that the hard band oscillation is systematically faster than, and hence 'runs away from', the soft band within a coherent interval, and thus that the two lightcurve possess a different QPO frequency. 
Furthermore, the phase lag resets at the start of each coherent interval, causing the QPO to stay coherent on longer timescales. This reset is already visible as the turnover around a cycle fraction of $0.9$, which is present in almost all observations. 
We will discuss the presence and interpretation of this turnover in more detail in the next section. 

In order to test our method for systematic effects and biases, we repeat the analysis, but randomize the measured start times of the coherent intervals. In other words, the assigned label of each segment is randomized,
 and does not respresent the actual segment position. We show the result of this check in panel 3 of Figure \ref{fig:main_h}. All phase lags are consistent with the grey band, which represents the average phase lag, 
independently calculated by averaging all CCFs without regard for their positions. This consistency indicates that the increase in panel 1 is inherently linked to the existence of and position within the coherent intervals. 

Panels 2 and 4 show the phase lag behaviour as a function of absolute position. In order to compare phase lag behaviour between coherent intervals of different lengths, we apply an extra selection before averaging the CCFs: segments
should not only have the same absolute position, but also be located in coherent intervals of the same total length. Interestingly, as can be seen in panel 2, shorter coherent intervals systematically show a steeper phase lag increase, but typically reach the same maximum lag before resetting. This suggests an underlying relation between the origin of the coherent intervals and the observed phase lag behaviour, that we will discuss in section \ref{sec:discussion}. 
Similar to panel 3, panel 4 shows the results of randomizing the assigned labels. Again, the observed effect disappears and only the average phase lag remains. 

In figure \ref{fig:main_l}, we show the same relations for a lower frequency QPO, where the frequency difference is opposite (i.e. the QPO frequency decreases with energy). The general trends are similar to, but, as expected, in the opposite direction as those in Figure \ref{fig:main_h}. The main difference is the presence of a non-zero phase lag at start of the coherent interval. The similarity of the phase lag behaviour at both high and low frequencies suggests that the runaway is a global characteristic of the observed QPO. 

\begin{figure*}
  \begin{center}
    \includegraphics[width=\textwidth]{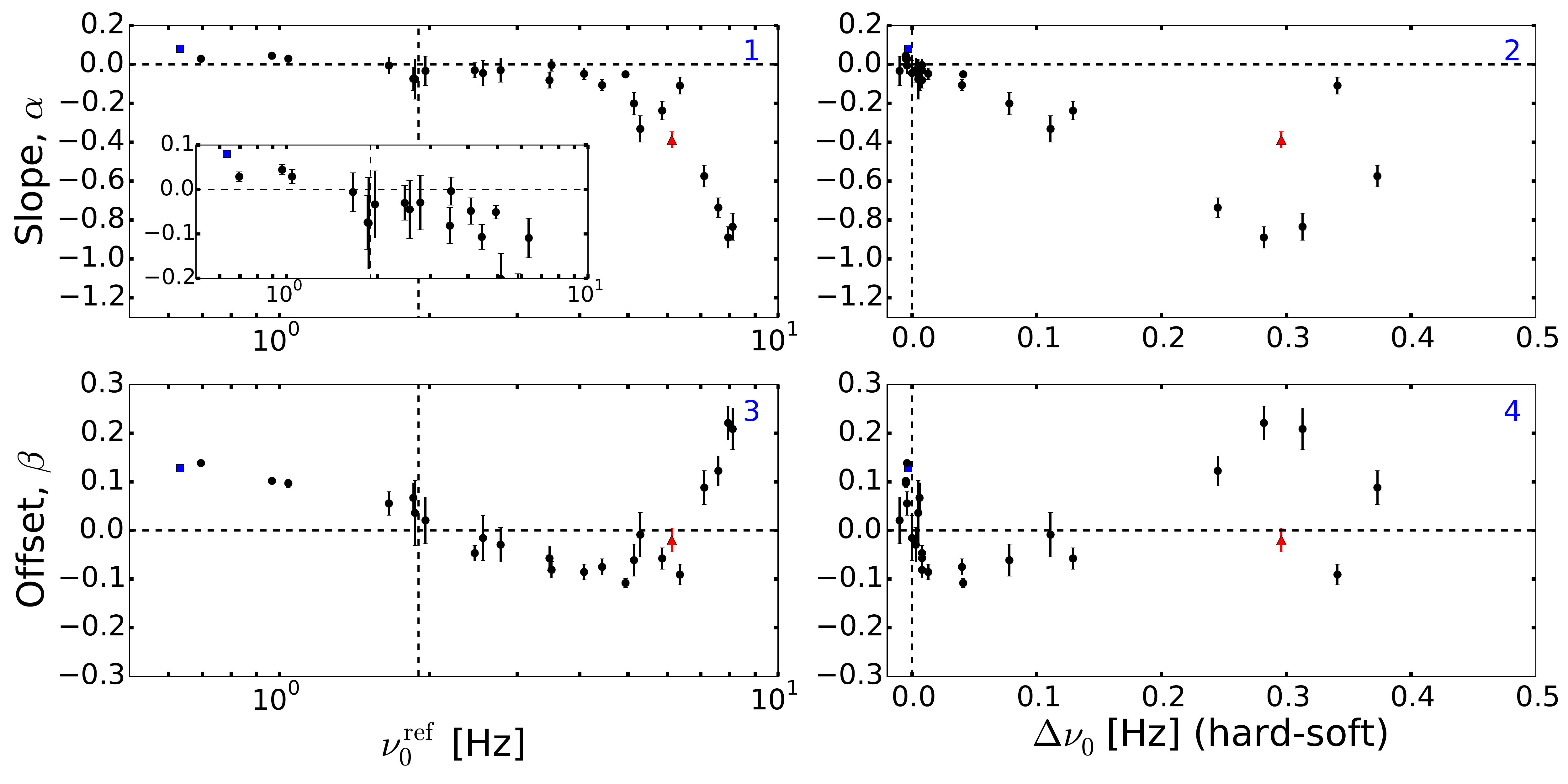}
    \caption{Fitted slope $\alpha$ and offset $\beta$ of the phase lag evolution plotted against both QPO frequency $\nu^{\rm ref}_{0}$ and frequency difference $\Delta \nu_0$. The vertical dotted line in panels 1 and 3 indicates $\nu^{\rm ref}_{0} = 1.9$ Hz, where the average phase lag at the QPO frequency changes sign. The red triangle and blue square correspond to the slope and offset of the phase lag evolution in panel 1 of Figures \ref{fig:main_h} and \ref{fig:main_l}, respectively. The inset in panel 1 shows a zoom in of the region below $5$ Hz, showing a significant change from positive to negative slopes around $1.9$ Hz. }
    \label{fig:all}   
  \end{center}
\end{figure*}

As a quantification of the increase in phase lag, we fit the phase lag as a function of fractional position with a straight line with non-zero offset:
\begin{equation}
\phi(x) = \alpha x + \beta
\label{eq:evolution}
\end{equation}
where $x$ is the fractional position. We do not propose this as the correct empirical representation of the lag behaviour, but it allows us to characterize
and compare all observations: the slope $\alpha$ quantifies the total change in phase lag in a full coherent interval and the offset $\beta$ estimates the starting phase lag. In other words, the slope characterizes the runaway between the two energy bands. Figure \ref{fig:all} summarizes all observations by plotting the slope and offset of this straight line model against both full-band QPO frequency $\nu^{\rm ref}_{0}$ and QPO frequency difference between the hard 
and soft band, $\Delta \nu_0$. The slope and offset corresponding to the first panel in Figures \ref{fig:main_h} and \ref{fig:main_l} are shown as the red triangle and blue square, respectively. Due to the relatively large uncertainties at the fractional position of $0.95$, we find no significant difference between fitting all phase lags or disregarding the turnover. 

Figure \ref{fig:all} shows that the runaway effect within coherent intervals is present in all observations with a non-zero frequency difference, and is small but signicant for small values of $\Delta \nu_0$. The sign of the slope, indicating which energy band
lags the other, is consistent within one sigma with the sign of $\Delta \nu_0$ in all observations. Panel 2 clearly shows that the slope increases as the frequency difference grows, as expected if the runaway is caused by a frequency difference. 
Panel 3 shows that the offset, i.e. the phase lag at the start of the coherent interval, follows a log-linear dependence on $\nu^{\rm ref}_{0}$, similar to the behaviour of the average phase lag in \citet{qu10}, \citet{pahari13} and \citet{reig00}, up to $\nu^{\rm ref}_{0} \sim 3.5$ Hz. At higher frequencies, the offset diverges from the log-linear relation back to positive values. The observations with these diverging offsets correspond to the observations with large frequency differences in panel 4. Using a completely different approach, \citet{qu10} concluded that the frequency difference contributes only a small fraction to the average phase lag at low frequency but becomes dominant above $\nu^{\rm ref}_{0} \sim 3.5$ Hz. In our analysis, this is visible in this divergence of $\beta$ from a log-linear dependency on frequency above approximately the same value of $3.5$ Hz. 

\section{Discussion and conclusion}
\label{sec:discussion}

\subsection{Robustness of the method}
\label{sec:robust}
We have tested our method for systematic biases at several important steps and have not found any significant effects. The choice of energy band to determine the start times of the coherent intervals has no effect on our results: our conclusions hold when using either the full, hard or soft band in this process. The same holds for the definition of a QPO cycle used while tracking the filtered lightcurves: we find no difference between defining a QPO cycle as two consecutive maxima or crossings of the mean photon rate. We find slight changes in our results if we replace the optimal filter by a tophat filter, which simply removes all variability outside a frequency range of $\nu_{\rm QPO} \pm \rm FWHM$. The phase lags do still increase in the coherent interval, but with a shallower slope. However, these changes can be expected since the tophat filter is less efficient than the optimal filter. This causes a decrease in the accuracy of our filtered QPO amplitude estimation and thus the determination of the coherent intervals. Finally, we find no significant changes in our results if we fit the CCFs with other functions, such as a damped sine wave, instead of a skewed Gaussian model. However, we apply the skewed Gaussian model to more accurately account for the asymmetry in the CCF when $\Delta \nu_0$ is significant. 

The robustness of our method also follows from the effect of randomizing the coherent interval start times (panels 3 and 4 in Figures \ref{fig:main_h} and \ref{fig:main_l}). In all observations, the phase lag evolution disappears and only the average phase lag remains. This shows that our method is capable of correctly estimating known phase lags. It also indicates that the observed phase lag evolution is inherently connected to the coherent intervals and cannot be attributed to biases in our method. Alternatively, if we distribute the start times evenly over the full lightcurve, the phase lag evolution disappears as well. This reinforces the notion that the evolution does not simply occur in any short time scale segment, but only in the actual coherent cycles. 

In order to interpret our results as a property of the QPO, we make the assumption that the phase lag at $\nu_{\rm QPO}$ is dominated by the QPO. For observations with a full-band QPO frequency $\gtrsim 5 Hz$, the lag spectrum shows a clear, distinct feature at the QPO frequency, while at other frequencies it appears featureless and hovers around zero. This implies that the phase lag is dominated by the QPO. At lower full-band QPO frequencies, these features in the lag spectrum become more diluted. Hence, we apply another test to our method: instead of calculating the CCF from optimally filtered lightcurves, we calculate it from lightcurves produced using a narrow tophat filter. If this narrow tophat filter is not centered on the QPO frequency, this method simply searches for phase lag evolution associated with the BBN. We perform this test several times per observation, sliding the tophat filter from low to high frequency. We find that the observed phase lag evolution is only present when applying the filter at the QPO frequency, and disappears otherwise. This indicates that the phase lag evolution is fundamentally linked to the QPO mechanism, and not to the BBN.

\subsection{Phase lags}

\begin{figure}
  \begin{center}
    \includegraphics[width=\columnwidth]{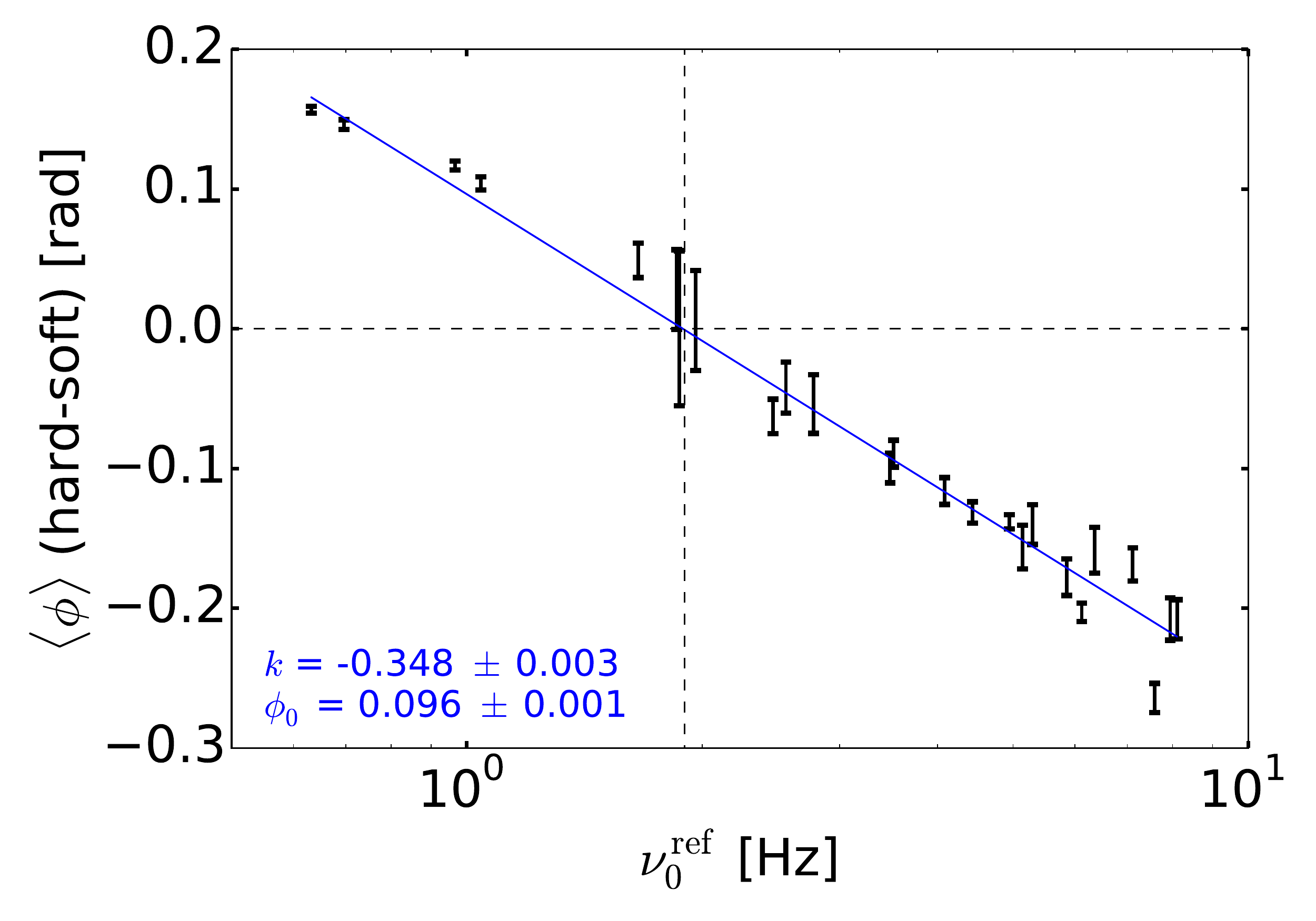}
    \caption{The average phase lag $\langle \phi \rangle$ as a function of fundamental QPO frequency $\nu^{\rm ref}_{0}$. Positve values of $\langle \phi \rangle$ correspond to hard lags. The straight line shows a log-linear model fit of $\langle \phi \rangle = k\log \nu^{\rm ref}_{0} + \phi_0$.}
    \label{fig:average_lags}   
  \end{center}
\end{figure}

As previously shown by \citet{qu10}, \citet{pahari13} and \citet{reig00}, the average phase lag measured for an entire observation evolves smoothly with the QPO frequency of that observation. We demonstrate this relation in Figure \ref{fig:average_lags}, where we plot average phase lag determined from the CCF versus QPO frequency for each analysed observation. Our results from the CCF confirm the aforementioned smooth evolution, which was previously only measured using the cross spectrum. To quantify this evolution, we fit the average phase lags with a simple log-linear model 
\begin{equation}
\langle \phi \rangle = k\log(\nu^{\rm ref}_{0}) + \phi_0 
\label{eq:loglin}
\end{equation}
which yields $\phi_0 = 0.096 \pm 0.001$ rad and $k = -0.348 \pm 0.003$ rad for $\nu^{\rm ref}_{0}$ in Hz. These values correspond to a switch from hard to soft lags at a QPO frequency of approximately $1.9$ Hz. Interestingly, this is consistent with the frequency where $\Delta \nu_0$ changes sign. 

It seems remarkable that the phase lags show such complexity within each coherent interval, yet the average phase lag falls on such a neat relation. An apparent discrepancy is clearly visible in Figure \ref{fig:all}. As long as the slope $\alpha$ is small ($\nu^{\rm ref}_{0} < 3.5$ Hz), the offset $\beta$ follows a log-linear relation with frequency. This is expected, since in the absence of any significant phase lag evolution, the offset will simply track the average phase difference. However, for higher QPO frequencies, where $\Delta \nu_0$ is large and the hard band quickly runs away from the soft band in each coherent interval ($\nu^{\rm ref}_{0} > 3.5$ Hz, $\alpha$ becomes signifcantly less than $0$), the offset $\beta$ diverges from the log-linear relation. This behaviour of the offset allows the average phase lags to remain on a log-linear relation with QPO frequency, by compensating for the short timescale phase lag evolution. 

\begin{figure}
  \begin{center}
    \includegraphics[width=\columnwidth]{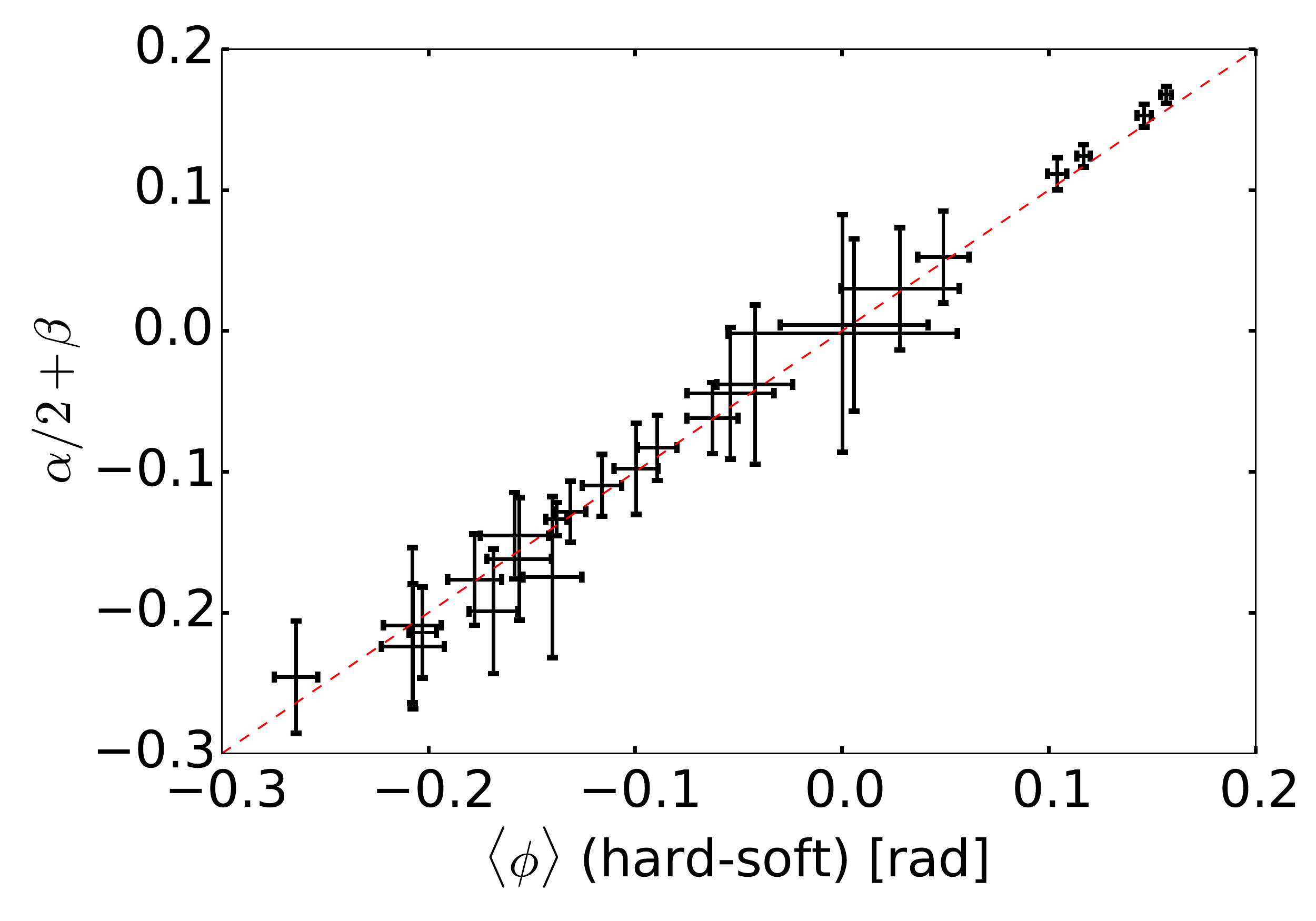}
    \caption{The measured average phase lag, $\langle \phi \rangle$, versus the predicted average phase lag based on the fitted phase lag evolution, $\alpha/2+\beta$. The dotted line shows the diagonal. The tight correlation indicates that the latter is an adequate predictor of the average phase lag at all QPO frequencies.}
    \label{fig:fundamental}   
  \end{center}
\end{figure}

To demonstrate this compensating behaviour, we plot the average phase lags measured from the CCF against those measured from the phase lag evolution (Equation \ref{eq:evolution}) in Figure \ref{fig:fundamental}. The latter are simply given by $\langle \phi \rangle = \langle \alpha x + \beta \rangle = \alpha/2 + \beta$. All observations clearly scatter around the dotted line, which represents the diagonal. Even at the most negative phase lags, corresponding to the highest QPO frequencies and largest values of the slope $\alpha$, the relation holds. The relatively large errorbars in the y-direction are caused by correlations between $\alpha$ and $\beta$ in fitting the phase lag evolution, which are not large enough to explain the relation over the entire range of observations. Thus, this tight agreement over the entire range indicates that the slope $\alpha$ and offset $\beta$ of each observation indeed compensate such that the average phase lag depends log-linearly on QPO frequency. This shows that the phase lag evolution simply pivots around the centre of each coherent interval. Of course, this does not explain why on long timescales, fundamentally, the average phase lag should follow a log-linear dependence on QPO frequency. 

A possible way to explain the smooth log-linear relation between the frequency and phase lag on long timescales, could be by considering the dependence of phase lag on photon energy. \citet{pahari13} studied this energy dependence and found that the phase lag at the QPO frequency follows a log-linear dependence on photon energy. A similar relation is found by \citet{qu10}, and hints of it are present in \citet{reig00}. The exact slope of this relation varies between different observations, and several observations also show hints of a break around $\sim 4-6$ keV. Next to GRS 1915+105, \citet{wijnands99} showed log-linear relations between phase lag and energy in XTE J1550-564. The smooth relation between phase lag and QPO frequency might be equivalent to a simple rotation of the dependence of phase lag on energy, where the rotational angle depends on the frequency of the QPO. This scenario is encouraged by the presence of the same phase lag-frequency relation in \citet{pahari13}, \citet{qu10} and our results, despite differences between the compared energy bands, both in width of the bands and centroid energies. In section \ref{sec:model}, we propose a possible mechanism that can explain these two observed dependencies of the phase lag and their possible equivalence based on a simple toy model.

\subsection{Decoherence of the QPO}

As part of our analysis, we have identified the envelopes in the QPO amplitude as so-called coherent intervals in all observations. These successive coherent intervals can be interpreted as subsequent, independent excitations of the QPO mechanism with a lifetime of approximately a coherence time. In this scenario, each single QPO excitation starts with the same initial phase lag between the hard and soft band. Subsequently, the frequency difference leads to the observed phase lag evolution as the excitation decays through the coherent interval. The observed turnover in the phase lag evolution could arise from the emergence of the next QPO excitation, which takes over from the previous one. In this interpretation, the frequency difference would play a vital role in the decay of the coherence through each interval. Here, we will discuss earlier evidence for separate QPO excitations and the relation between QPO frequency difference and (de)coherence. In section \ref{sec:model}, we will discuss our results in the light of a simple physical toy model. 

Using adaptive decomposition in a single $\sim 4$ Hz QPO observation of XTE J1550-564, \citet{su15} identify \textit{intermittent oscillations} of $\sim 3$ seconds, where the QPO is coherent. Similar to our coherent intervals, these intermittent oscillations consist of high amplitude segments of the QPO lightcurve and are separated by short, lower amplitude segments. \citet{su15} interpret this behaviour within the Lense-Thirring precession model as the alternation between steady precession (high amplitude) and unstable precession (low amplitude) of the inner hot flow. The $\sim 3$ second timescale is interpreted as the the viscous timescale at the outer radius of the inner hot flow. BHB XTE J1550-564 shows energy-dependent QPO frequencies similar to GRS 1915+105 \citep{li13}, suggesting that these intermittent oscillations are in fact the same as coherent intervals.  

\citet{lachowicz10} expanded the standard Fourier methods using wavelet analysis and the Matching Pursuit algorithm to track the QPO properties over time. They are able to reconstruct the observed properties of the QPO using multiple short timescale QPO signals with random amplitudes. Given these existing results in XTE J1550-564, the similarity between intermittent oscillations and coherent intervals, and the simple explanation of the short timescale phase lag evolution, we adopt the interpretation of coherent intervals as a manifestation of distinct, independent excitations of the QPO mechanism. 

Identifying coherent intervals as separate QPO excitations does not provide an explanation for the decoherence of the QPO through the interval. However, the systematic phase lag evolution might be accountable for that. Interestingly, we find that the speed of the phase lag evolution is linked to the length of the coherent interval: as is visible in panel 2 in Figures \ref{fig:main_h} and \ref{fig:main_l}, shorter coherent intervals have a significantly quicker phase lag evolution. Moreover, for all coherence lengths, the phase lag reaches approximately the same maximum value for the same QPO frequency before turning over. Only when we account for the length of the coherent interval, are all phase lag evolutions mapped onto one relation (panel 1 in Figures \ref{fig:main_h} and \ref{fig:main_l}). In other words: the length of the coherent interval appears to be set at the start of each interval and the phase lag evolves accordingly. This suggests a causal relation between the length of a coherent interval and the frequency difference in that particular coherent interval: a larger frequency difference leads more quickly to a large phase lag, which in turn reduces the coherence of the QPO and thus the length of the coherent interval. This is consistent with the interpretation of coherent intervals as independent QPO excitations, since in this case each individual excitation can have a distinct frequency difference. Even though the hard and soft band are not completely out of phase at the end of each coherent interval, the phase lag could still contribute to the decoherence.  

\begin{figure}
  \begin{center}
    \includegraphics[width=\columnwidth]{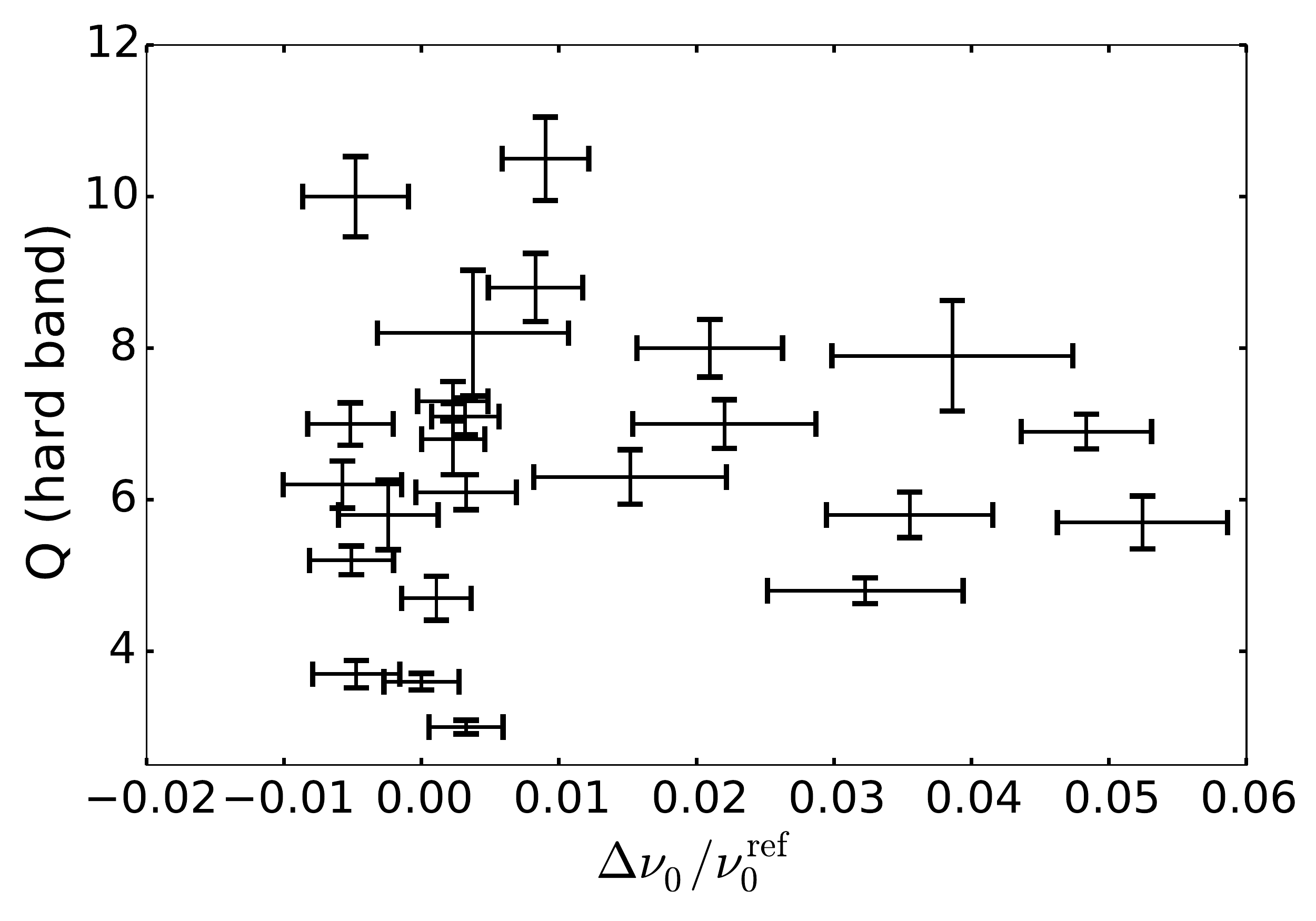}
    \caption{The quality factor $Q$ of the QPO in the hard band of each observation as a function of the relative QPO frequency difference $\Delta \nu_0/\nu_0^{\rm ref}$.}
    \label{fig:Q}   
  \end{center}
\end{figure}

A final property of the coherent intervals is not directly accounted for by the observed phase lag evolution: the QPO shows coherent intervals in all observations, even when no phase lag evolution is observed. In other words, the oscillations are always \textit{quasi}-periodic, regardless of the presence of a frequency difference between energy bands. This can be seen clearly in Figure \ref{fig:Q}, where we plot the Q-value of the hard band as a function of relative frequency difference $\Delta \nu_0 / \nu^{\rm ref}_{0}$. In the Figure, no clear relation between $\Delta \nu_0$ and $Q$ exists, contrary to what is expected if the frequency difference and resulting phase lag evolution causes the decoherence. In the next section, we offer a toy model based on our results that accounts for this last difficulty in explaining the decoherence through the phase lag evolution. 

\subsection{A unifying model: differential precession and spectral evolution}
\label{sec:model}
Any model to explain the decoherence of QPO excitations should account for both the sign of the frequency difference at low QPO frequencies and the link between phase lag evolution, QPO frequency difference and length of the coherence interval. The former is especially challenging: phenomenologically, an increase of frequency with energy can be understood by considering different radii in the accretion flow. If the QPO frequency can be associated with characteristic timescales in the accretion flow, a higher frequency would correspond to a smaller radius. Simultaneously, from these smaller radii, a harder spectrum is expected to be emitted. In this simple picture, it is possible to account for a positive correlation between frequency and energy. However, this simple scenario is unable to explain the anti-correlation between frequency and energy that is observed at low QPO frequencies. This does not even touch upon the fact that the slope of the relation between energy and QPO frequency changes very systematically with $\nu^{\rm ref}_{0}$, which adds another layer of difficulty to this problem. Since we conclude that the frequency difference is intrinsic, any viable interpretation should relax either the assumption that a smaller radius corresponds to shorter timescales, or to a harder emission. Moreover, any model should also account for the observed properties of the phase lags on long timescales, both as a function of frequency and energy. 

One possible interpretation of our results and the observed sign of $\Delta \nu_0$ incorporates spectral changes of the inner accretion flow as the full-band QPO frequency decreases. In this approach, we stick to the anticorrelation between oscillation frequency and radius and instead loosen the view that hardness tracks proximity to the black hole. As the QPO amplitude is known to depend on inclination, favoring a geometric QPO mechanism \citep{motta15,heil15}, we assume a toy model where the QPO is caused by vertical precession of the inner flow \citep[as in e.g.][]{ingram09}. In this toy model, we consider this inner flow to consist of two separate halves, which differentially precess at slightly different frequencies. In agreement with the expected dependence of frequency on radius, we assume that the inner half \textit{always} causes a higher QPO frequency than the outer half: $\Delta \nu_0 (\rm Inner - Outer) > 0$. Dividing the inner flow into two halves is of course an extremely simplified picture. More realistically, the inner flow might show a more continuous evolution in precession frequency over radius. However, as our analysis only compares two broad energy bands, the simplified approach is sufficient for a quantitative comparison with our results. 

\begin{figure}
  \begin{center}
    \includegraphics[width=\columnwidth]{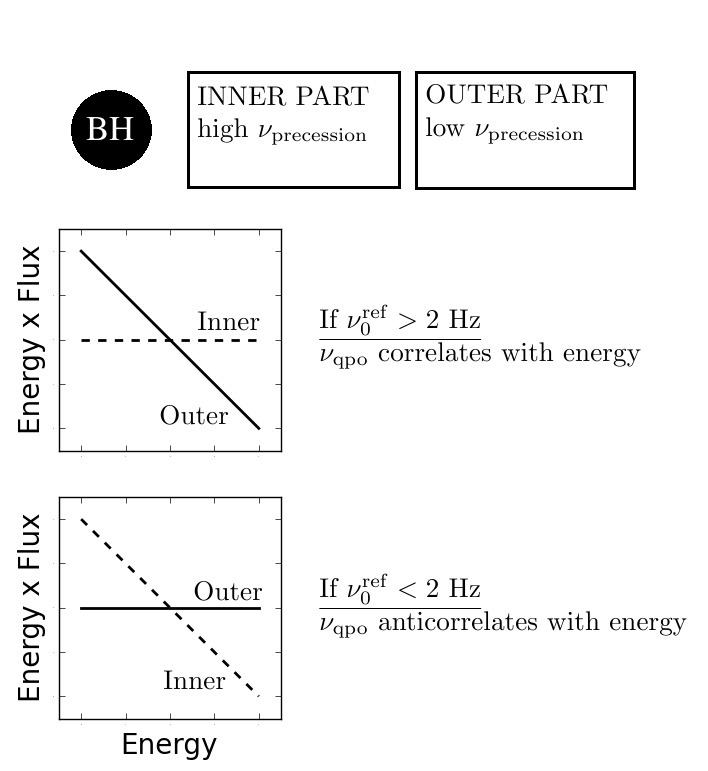}
    \caption{Cartoon depiction of the proposed geometry and spectral evolution in the differential precession model. Combined, the inner and outer part constitute the inner accretion flow, while the disk is not shown. $\nu_0^{\rm ref}$ corresponds to the QPO frequency in the full energy band, while $\nu_{\rm qpo}$ refers to the QPO frequency observed in a narrow energy band.}
    \label{fig:sketch}   
  \end{center}
\end{figure}

To qualitatively explain our results in the light of this simplified model, we introduce an evolution of the spectral properties of the inner and outer half of this differentially precessing flow. At high full-band QPO frequencies ($> 2$ Hz), the inner half has a harder spectrum than the outer half, as expected. As the full-band frequency decreases, the spectral shapes change, such that at $\sim 2$ Hz, both the inner and outer spectrum have the same hardness. Below $\sim 2$ Hz, the spectral hardness of the inner and outer parts have switched around so that the outer half is harder than the inner half. In this scenario, the hard and soft band are not always accurate tracers of the inner and outer parts of the inner flow. This causes the observed $\Delta \nu_0 (\rm hard - soft)$ to change from positive to negative as frequency evolves from high to low, while the physical frequency difference $\Delta \nu_0 (\rm Inner - Outer)$ stays positive, as expected. In Figure \ref{fig:sketch}, we show a simplified sketch of the model geometry and associated spectral evolution. 

This differential precession model is able to explain several of the observed features of the QPO frequency and phase lag. A first advantage is the fact that the difference in QPO frequency is always intrinsically present, even when it is not observed. Thus, the phase lag between the inner and outer half can always evolve and decohere the precession. As each QPO excitation $\Delta \nu_0$ can change stochastically, coherent intervals will vary in length. Thus, this mechanism can explain both why the QPO is always quasi-periodic, even when no frequency difference is observed, and why the phase lags evolve quicker as the QPO decoheres faster. While intrinsically, the inner and outer half might get completely out of phase as the QPO decoheres ($\phi_{\rm intrinsic} = \pm \pi$), we observe the spectrally weighted energy bands. Thus, the maximum observed phase lags are actually smaller than $\phi = \pm \pi$, as in our results. 

Secondly, this model can explain the observed properties of the average phase lags as a function of both energy and frequency. On average, the outer half will lag behind the inner half with a lag of $\sim \pi/2$, assuming that the two halves get completely out of phase during a coherent interval. When the inner half has a harder spectrum (high full-band QPO frequencies), this causes an observed soft lag. However, when the spectrum is reversed (low full-band QPO frequencies), the observed phase lag becomes hard. Thus, this model can explain the change from hard to soft lags as the QPO frequency increases. This intrinsically constant average phase lag will also display an energy dependence similar to the QPO frequency, as is observed. An important argument for our interpretation is the peculiarity that the energy dependence of both the QPO frequency and phase lag flip over at $\sim 2$ Hz, where the lags change from hard to soft. In this model, this frequency of $\sim 2$ Hz is simply the frequency where the spectra of the inner and outer halves have the same shape. 

This interpretation with a differentially precessing inner flow introduces a fairly complicated extra geometry, with different intrinsic precession frequencies. \citet{nixon12} have shown that in cases with a sufficient misalignment between the outer disk and the black holes spin axis, Lense-Thirring torques can break the inner flow into separately rotating rings. However, to explain the coherent intervals as subsequent QPO excitations, these rings should set up and decay on very short timescales. Thus, it could be envisioned more as the introduction of a gradual warp in the inner accretion flow. However, due to our limited energy resolution of two broad energy bands, it is not possible to distinguish between a gradual warp or two independently precessing parts. 

Several possible explanations for the required spectral evolution could exist. Here, we discuss two options, although others might be possible as well. First, the reflected spectrum could contribute: in the differential precession model, this reflected spetrum is expected to be dominated by photons emitted by the outer precessing half, as this blocks photons from the inner half. Thus, the reflected spectrum will show a relatively low precession frequency. However, the hardness of the reflected spectrum can change: based on observed relations between spectral and timing properties of accreting black holes, the disk ionisation is expected to increase with full-band QPO frequency (see e.g. \citealt{gilfanov10} for an overview or \citealt{shaposhnikov09} and \citealt{garcia15} for specific examples). This would imply a relatively soft reflected spectrum at high full-band QPO frequencies, and a relatively hard reflected spectrum at low full-band QPO frequencies. As the reflected spectrum shows a slow precession, this effect could possibly cause the QPO frequency to change from increasing to decreasing as a function of energy.

A second explanation could be the presence of an extra cooling process very close to the black hole. At high full-band QPO frequency, the differentially precessing inner flow is expected to be small. In this scenario, the cooling process would be present in the entire flow, leading to a steepening of the spectrum everywhere. At lower full-band QPO frequencies, the inner flow extends further out - if the extra cooling term is only present in the innermost part, this might cause a steepening of the spectrum only for the regions precessing most quickly. Thus, at these low frequencies, this additional cooling term could cause the outer part to be harder than the inner part. Such an extra cooling process could for instance be pair-production and annihilation, which, if present, would be expected close to the black hole due to the high densities in the flow \citep{fabian15}.

The origin of the spectral evolution as a function of full-band QPO frequency might also account for the differences in energy dependence of the QPO frequency between GRS 1915+105, XTE J1550-564 and H1743-322, and the lack of observed energy dependence in other BHBs. For example, if the reflected spectrum plays a significant role, its inclination-dependence could contribute to differences in behaviour between these sources. The role of reflection could be tested by considering the energy-dependence of the QPO frequency below $\sim 2$ keV: if this role is indeed significant, reprocessing of photons by the disk is also expected at lower energies. As the reprocessed photons also originate from the slower oscillating outer half, this would cause a low observed QPO frequency below $\sim 2$ keV, similar to the QPO frequency in our hard band. This energy range could in the future be probed by for instance \textit{the Neutron star Interior Composition ExploreR} \citep[NICER;][]{gendreau12} to search for this effect.

\section{Conclusion}

We have developed and implemented a new model-independent method to compare QPO properties between different energy bands on short timescales. This has allowed us to test whether the observed energy dependence of the QPO frequency corresponds to a genuine difference in this frequency between energy bands. We find that no clear (anti)correlations exist between QPO frequency and amplitude in filtered QPO lightcurves, and that the phase lag between two energy bands increases throughout each coherent interval ($\sim 5-10$ QPO cycles), before resetting at the end. We also find that the speed of this phase lag evolution is larger when the coherent interval is shorter. These results lead us to conclude that the QPO possesses a genuinely different frequency in different energy bands. We interpret this in the context of a geometric toy model, where the QPO is caused by differential precession of the innermost accretion flow. This toy model allows us to qualitatively explain (1) the energy-dependence of the QPO frequency and phase lag, (2) the smooth relation between full-band QPO frequency and phase lag and (3) the decoherence of the QPO on the timescale of $\sim 5-10$ QPO cycles.

Our newly developed method to track the evolution of QPO properties on short timescales shows promise for a more general \textit{coherent interval resolved} analysis of the QPO. For instance, next to the phase lag, it is possible to track the QPO frequency and frequency difference and compare these within and between coherent intervals. This would yield a much more detailed look at the relations between the length of coherent intervals, the QPO frequency difference and the speed of the phase lag evolution. Such an analysis could shed more light on the intricacies of the underlying QPO mechanism. Applying this method using narrower energy bands could also provide a useful handle on the physical location where the phase lag evolution and the frequency difference originate. Finally, the method could be extended by implementing the Hilbert-Huang transform (\citealt{huang98}, see \citealt{su15} for its application to QPO lightcurves) to obtain the instantaneous frequency and amplitude in each position in the filtered lightcurve. This would increase the accuracy of the determination of the coherent intervals, which would be especially useful when applying this technique to sources with a lower X-ray flux. 

\section*{Acknowledgements}

We thank the referee for insightful comments on this work. We also thank Michiel van der Klis for helpful discussions on the interpretation of these results. This research has made use of data obtained through the High Energy Astrophysics Science Archive Research Center Online Service, provided by the NASA/Goddard Space Flight Center. AI acknowledges support from the Netherlands Organization for Scientific Research (NWO) Veni Fellowship, grant number 639.041.437.




\bibliographystyle{mnras}







\bsp	
\label{lastpage}
\end{document}